\documentclass{appolb}
\usepackage{epsfig}

\newcommand{\eq}{\begin{equation}}
\newcommand{\eqx}{\end{equation}}
\newcommand{\eqn}{\begin{eqnarray}}
\newcommand{\eqnx}{\end{eqnarray}}
\newcommand{\f}[2]{\frac{#1}{#2}}

\newcommand{\lra}{\longrightarrow}

\renewcommand{\th}{\theta}
\newcommand{\sg}{\sigma}
\newcommand{\al}{\alpha}
\newcommand{\alef}{\alpha'_{eff}}

\newcommand{\dl}{\delta}

\newcommand{\DD}{{\cal D}}
\newcommand{\AAA}{{\cal A}}

\newcommand{\qb}{\bar{q}}
\newcommand{\cor}[1]{\left\langle{#1}\right\rangle}

\newcommand{\rr}[4]{#1, {\it #2 \/}{\bf #3} #4}

\title{QCD, diffraction and string theory}

\author{Romuald A. Janik
\thanks{e-mail:{\tt janik@nbi.dk}.
Presented at the 10th International Workshop on Deep Inelastic
Scattering, DIS2002, Krakow, Poland, 30 April--4 May 2002.}
\address{The Niels Bohr Institute,\\
Blegdamsvej 17, DK-2100 Copenhagen,\\ 
Denmark\\
and\\
M. Smoluchowski Institute of Physics,\\ 
Jagellonian University,\\
Reymonta 4, 30-059 Cracow, \\
Poland}
}

\begin{document}

\maketitle

\begin{abstract}
Recently, string theory on some specific curved backgroud 
spacetime geometries has been conjectured to be equivalent
to certain gauge theories (AdS/CFT correspondence).
This correspondence may be used to investigate 
the non-perturbative regime of gauge theories.
I describe its application to the study of soft scattering 
amplitudes in a confining gauge theory.
I describe two qualitatively different applications:
amplitudes with vacuum quantum number exchange (Pomeron-like),
amplitudes with Reggeon exchange.
The last case requires going beyond eikonal approximation on
the gauge theory side.
\end{abstract}

\section{Introduction}

In the phenomenological description of (soft) diffractive 
processes in the Regge limit
a prominent role is played by the various Regge poles and their
couplings, like e.g. the 3-Pomeron vertex. Experimentally their
properties are well established. One has two distinct
cases. The dominant trajectory with vacuum quantum number exchange is
the Pomeron, leading to amplitudes behaving like $s^{1.08+0.25t}$ in
the soft regime. The other family of trajectories correspond to
Reggeon exchanges (mesonic trajectories) and involves typically the
exchange of flavour. The amplitudes behave here quite differently ---
with the leading amplitudes like $s^{0.55+1t}$. In particular the
slope is almost exactly {\em four} times larger than for the Pomeron.

It remains a formidable challenge to understand/derive these
properties from more fundamental principles. The major stumbling block
is of course the inherently nonperturbative character of these processes.
In this talk I will describe an approach \cite{us2,fluct,usr} to
calculating the properties of these trajectories\footnote{For other
approaches to the soft Pomeron see \cite{SZN,KKL}.} within the framework
of the AdS/CFT correspondence. 

The AdS/CFT correspondence \cite{ma98} is the conjectured equivalence between
certain gauge theories and string theories on appropriate curved
backgrounds. The utility of the correspondence comes from the fact
that strong coupling problems in gauge theory side are mapped to
quasi-classical problems on the string theory side. A precise
version of this correspondence does not exist so far for QCD, so we
used a generic version for a theory with confinement.

\section{Pomeron dominated amplitudes}

In the Pomeron channel, since we want to study soft processes
and no flavour
quantum numbers are exchanged it suffices to use the eikonal
approximation \cite{Nacht}. In this approximation the impact
parameter $q\qb$ scattering 
amplitude is given by a correlation function of two Wilson lines which
follow classical straight line trajectories:
\eq
\label{e.lines}
A(s,L)=i s \cor{e^{i\int_{L_1} A} e^{i\int_{L_2} A}}
\eqx

Technically we performed the calculation of the
Wilson line correlator in Euclidean space (using AdS/CFT
correspondence) as a function of the impact parameter $L$
and the relative angle $\th$ between the two lines.
The result $A(\th,L)$ was then continued back to {\em Minkowski} space
using the substitutions $\th \lra -i\chi \sim -i \log s$ and $T\to i T$.
The above procedure was first used within the eikonal approximation in
perturbative QED and QCD in \cite{Megg}. 

The $q\qb$ amplitude as it stands is IR divergent. We regularized it by
introducing a temporal cut-off by taking the Wilson lines to be of finite
length, and adding gauge `connectors' at both ends to close the lines
into a loop.

Within the AdS/CFT correspondence the expectation value of a Wilson
loop at strong coupling is given by \cite{loopsads}
\eq
\cor{W(C)} \sim Fluctuations(\Sigma_{minimal}) \cdot e^{-\f{1}{2\pi \al'}
Area(\Sigma_{minimal})} 
\eqx
where $\Sigma_{minimal}$ is the surface of minimal area in the bulk of
the geometry which is spanned on the contour $C$, and the prefactor
represents the contrbution of quadratic fluctuations of the string
worldsheet around the minimal surface.
 
In our case, in the
confining regime the relevant surface will be (a sector of) a helicoid. 
The explicit formula for its area is \cite{us2}
\eq
\label{e.area}
Area= \int_{-L/2}^{L/2} d\sg 
\, \left\{T \sqrt{1+p^2 T^2}+ \f{1}{p}\log\left( p T +
\sqrt{1+p^2 T^2} \right) \right\}
\eqx
This result has to be analytically continued
to Minkowski space. Naively we would obtain a pure phase. 
However, due to the presence of logarithmic cuts in the complex plane,
we obtain very specific contributions when going through a cut. 
It is therefore interesting to explore the physical consequences of
these contributions. Consequently we have to perform
the substitution $\log(\ldots) \lra \log(\ldots)- 2\pi i n$,
with $n$ being some integer number.
Under this transformation, the amplitude gets a contribution:
\eq
e^{\f{1}{2\pi \alef} \f{L^2}{\th} 2\pi i n} \lra 
e^{-\f{1}{\alef} \f{L^2}{\log s} n}
\eqx
which is {\em independent} of the IR cut-off $T$. In the following we
will neglect the $T$ dependent terms assuming that $T$ is small (some
justification for this assumption was given in \cite{us2}). After Fourier
transform we obtain an inelastic amplitude with a linear Regge
trajectory:
\eq
(prefactor) \cdot s^{1+\f{\alef}{4} t}
\eqx
The prefactor here includes a $\log s$, further such contributions may come
from $\al'$ corrections. In the following we concentrate on the
dominant terms which give rise to a power-like $s^\al$ behaviour.

An interesting feature of this result is that the linear slope
$\alef/4$ characteristic of soft Pomeron exchange arose
through the analytic structure of the helicoid area.

The contribution of quadratic fluctuations was evaluated in
\cite{fluct} using the fact that the dual string theory in the AdS/CFT
picture is {\em critical}.
The piece that dominates after continuation to
Minkowski space for high energies is 
\eq
\label{e.fluct}
Fluctuations=
\exp \left( n_\perp \cdot \f{\pi}{24} \cdot
\f{\th}{2 \log\left( pT +\sqrt{1+p^2T^2} \right)} \right)
\ .
\eqx
Let us now perform the same analytical continuation to Minkowski
space, keeping in mind the substitution $log \to log -2\pi i n$. 
Furthermore we neglect the logarithmic $T$
dependent terms. The outcome is
\eq
\label{e.fluctsp}
Fluctuations=e^{\f{n_\perp}{96}\log s}=s^{\f{n_\perp}{96}}
\eqx

Putting together the above results we obtain finally for the trajectory
\eq
(prefactors)\cdot  s^{1+\f{n_\perp}{96}+\f{\alef}{4} t}
\eqx
The values for $n_\perp=7$ suggested by the AdS/CFT correspondence
\cite{kinar} would
give an intercept of $1.073$ (or 1.083 for $n_\perp=8$), very close to
the observed soft Pomeron intercept of 1.08. The phenomenological
value of $\alef$ extracted from the static quark-antiquark potential
is $\alef\sim 0.9 \, GeV^{-2}$. This gives the slope $0.225$ in
comparison with the observed one of $0.25$.

\section{Reggeon dominated amplitudes}

In order to be able to isolate an amplitude where the dominant
contribution will be given by a Reggeon exchange, we have to consider
a mesonic scattering amplitude with an {\em exchange} of two quarks
(see \cite{usr} for a detailed discussion). 
The (position-space) amplitude  typically involves
four fermionic propagators, two of which can be calculated in the
eikonal approximation (the spectator quarks), while the exchanged ones
are described within the worldline formalism:
\eq
S(x,y | \AAA)=\int \DD x^\mu(\tau) \, e^{-m \cdot Length} \cdot \left\{
\mbox{\rm Spin Factor} \right\} \cdot e^{i\int_{trajectory} A}
\eqx 
where the $\left\{\mbox{\rm Spin Factor} \right\}$ keeps track of the
spin 1/2 nature of the quarks. In the above expression the colour and
spin parts do 
factorize, which is very convenient for calculations using various
models of the nonperturbative gluonic vacuum.

The amplitude then becomes a path integral over the exchanged quark
trajectories, the integrand being the spin factors and a Wilson loop
formed by the spectators and the exchanged quarks \cite{usr}. 
In order to perform
the Wilson loop average we use the AdS/CFT correspondence. 
In the approximation when the exchanged quarks are light, one can
assume that the dominant contribution will come from the helicoid
minimal surface spanned by the spectator quarks whose upper and lower
boundaries are formed by the exchanged quark trajectories.
At this stage the expression for the amplitude has the following
structure (in the $m\to 0 $ limit):
\eq
is \int \DD x(\tau) \left\{\mbox{\rm Spin Factor} \right\} \cdot
e^{-S_{area}[x(\tau)]} 
\eqx
The $\left\{\mbox{\rm Spin Factor} \right\}$ essentially gives a $1/s$
suppression characteristic of an exchange of two spin $\f{1}{2}$ particles.
The form of the effective action $S_{area}[x(\tau)]$ for the quark
trajectories 
follows from the helicoid geometry. We evaluated the path integral
using the saddle point configuration $\dl S_{area}[x(\tau)]/\dl
x(\tau')=0$. The saddle point turned out to be imaginary,
consistent with an inelastic aplitude, and gave a linear trajectory
with a slope of $\alef$, exactly four times larger as in the pomeron
case. The final result including the fluctuations of the string
worldsheet around the helicoid gives the following result:
\eq
s^{0+\f{n_\perp}{24}+\alef t}
\eqx

\section{Discussion}

The AdS/CFT correspondence lays down a framework for addressing
nonperturbative properties of gauge theories including those
intrinsically linked with confinement. Linear trajectories with the
experimentally observed slopes arise naturally, although they are
encoded in analytical properties of the helicoid minimal
surface. String fluctuations lead to a rise of the intercepts. 
We observed a surprising quantitative agreement with the
experimentally observed soft Pomeron intercept, while the  
Reggeon intercept, while qualitatively reasonable is lower than the
observed value. It would be intersting to understand this better.

\bigskip

\noindent{\bf Acknowledgments.} The vast majority of the results
reported here were obtained in collaboration with Robi Peschanski.
This work was supported in part by KBN grant 2P03B09622 (2002-2004).

\end{document}